\documentclass[ijms,article,accept,pdftex,moreauthors]{Definitions/mdpi}

\firstpage{1}
\pubvolume{25}
\issuenum{14}
\articlenumber{7877}
\pubyear{2024}
\copyrightyear{2024}
\externaleditor{Academic Editor: Michal\linebreak Cichomski}
\datereceived{3 July 2024}
\daterevised{15 July 2024} 
\dateaccepted{16 July 2024}
\datepublished{18 July 2024}
\hreflink{https://doi.org/10.3390/\linebreak ijms25147877} 

\Title{Temporal Evolution of Defects and Related Electric Properties in He-Irradiated YBa$_{2}$Cu$_{3}$O$_{7-\delta}$ Thin Films}

\TitleCitation{Temporal Evolution of Defects and Related Electric Properties in He-Irradiated YBa$_{2}$Cu$_{3}$O$_{7-\delta}$ Thin Films}

\Author{{Sandra Keppert} $^{1}$\href{https://orcid.org/0009-0007-2609-143X}{\orcidicon}, Bernd Aichner $^{2}$\href{https://orcid.org/0000-0002-9976-9877}{\orcidicon}, Philip Rohringer $^{2,\dagger}$, Marius-Aurel Bodea  $^{1,\ddagger}$, Benedikt M\"uller $^3$,\linebreak Max Karrer $^3$, Reinhold Kleiner $^3$\href{https://orcid.org/0000-0002-2819-9774}{\orcidicon}, Edward Goldobin $^3$\href{https://orcid.org/0009-0005-6543-3653}{\orcidicon}, Dieter Koelle $^3$\href{https://orcid.org/0000-0003-3948-2433}{\orcidicon}, Johannes D. Pedarnig~$^{1}$\href{https://orcid.org/0000-0002-7842-3922}{\orcidicon}\linebreak and Wolfgang Lang $^{2,}$*\href{https://orcid.org/0000-0001-8722-2674}{\orcidicon}}

\AuthorNames{Sandra Keppert, Bernd Aichner, Philip Rohringer, Marius-Aurel Bodea, Benedikt M\"uller, Max Karrer, Reinhold Kleiner,  Edward Goldobin, Dieter Koelle, Johannes D. Pedarnig, and Wolfgang Lang}

\AuthorCitation{{Keppert, S.;} 
Aichner, B.; Rohringer, P.; Bodea, M.-A.; M\"uller, B.; Karrer, M.; Kleiner, R.; Goldobin, E.; Koelle, D.; Pedarnig, J.D.;~et~al.}

\address{%
$^{1}$ \quad Institute of Applied Physics, Johannes Kepler University Linz, {4040 Linz, Austria;} sandra.keppert@jku.at~(S.K.);~johannes.pedarnig@jku.at (J.D.P.)\\
$^{2}$ \quad Faculty of Physics, University of Vienna, {1090 Vienna, Austria}; bernd.aichner@univie.ac.at (B.A.) \\
$^{3}$ \quad  Physikalisches Institut, Center for Quantum Science (CQ) and LISA{$^+$}, 
University of T\"ubingen, {72076~T\"ubingen, Germany}; { \mbox{kleiner@uni-tuebingen.de (R.K.)}; gold@uni-tuebingen.de (E.G.)}; koelle@uni-tuebingen.de (D.K.)
}

\corres{Correspondence: wolfgang.lang@univie.ac.at}

\firstnote{Current address: Austrian Patent Office, 1200 Vienna, Austria.}
\secondnote{Current address: Infineon Technologies Austria AG, 9500 Villach, Austria.}

\abstract{Thin films of the superconductor YBa$_2$Cu$_3$O$_{7-\delta}$ (YBCO) were modified by low-energy light-ion irradiation employing collimated or focused He$^+$ beams, and the long-term stability of irradiation-induced defects was investigated. For films irradiated with collimated beams, the resistance was measured in situ during and after irradiation and analyzed using a phenomenological model. The formation and stability of irradiation-induced defects are highly influenced by temperature. Thermal annealing experiments conducted in an Ar atmosphere at various temperatures demonstrated a decrease in resistivity and allowed us to determine diffusion coefficients and the activation energy $\Delta E = (0.31 \pm 0.03)$ eV for diffusive oxygen rearrangement within the YBCO unit cell basal plane. Additionally, thin YBCO films, nanostructured by focused He$^+$-beam irradiation into vortex pinning arrays, displayed significant commensurability effects in magnetic fields. Despite the strong modulation of defect densities in these pinning arrays, oxygen diffusion during room-temperature annealing over almost six years did not compromise the signatures of vortex matching, which remained precisely at their magnetic fields predicted by the pattern geometry. Moreover, the critical current increased substantially within the entire magnetic field range after long-term storage in dry air. These findings underscore the potential of ion irradiation in tailoring the superconducting properties of thin YBCO films.}


\begin{document}


\section{Introduction}
{High-temperature} superconductors (HTSCs) are fascinating and are studied intensively due to their unconventional superconductivity. Among~the HTSCs, YBa$_2$Cu$_3$O$_{7-\delta}$ (YBCO) is a popular choice due to its easy-to-fabricate and non-toxic nature, as~well as its high critical temperature above the boiling point of liquid nitrogen. Various experiments have been conducted using masked or focused ion irradiation to modify the superconducting properties of YBCO and observe novel phenomena. These experiments include the creation of Josephson junctions~\cite{KAHL98,KATZ98,BERG05,CYBA15,MULL19}, nanoscale superconducting quantum interference devices~\cite{BERG06,LI20}, superconducting quantum interference filters~\cite{OUAN16}, and~the manipulation of Abrikosov vortex behavior through the fabrication of artificial periodic columnar defects, which serve as pinning centers for vortices~\cite{SWIE12,TRAS14,HAAG14,ZECH18a,AICH19,KARR24P} due to a local suppression of superconductivity~\cite{LESU93}.

Ion-beam structuring of YBCO using light elements such as He$^+$ ions has advantages over wet-chemical lithography and conventional ion-beam milling using Ga. Heavy ions with moderate energy are unsuitable as they have a limited penetration depth and can implant even into thin films, while swift heavy ions produce randomly distributed amorphous channels in YBCO~\cite{CIVA97R}. Irradiation with electrons~\cite{TOLP96} and protons~\cite{MEYE92} would require an impractically high dose per defect column to achieve a significant pinning effect. In~contrast, He$^+$ ions with energies of at least 30 keV can penetrate through thin YBCO films and create columns of point defects with reasonable doses. Furthermore, the~minimal lateral straggle of their trajectory makes them an ideal choice~\cite{LANG09,KARR24P}.

Thorough investigations on the long-term stability of ion irradiation-induced modification of YBCO are essential for potential technical applications of the modified material. However, there is a scarcity of data on the temporal evolution of electrical properties immediately following ion irradiation and after long-term storage at ambient conditions. Research has been conducted on the time-dependent properties of Josephson junctions created by focused He$^+$-ion-beam (He-FIB) irradiation~\cite{KARR24}, along with the effects of post-annealing in an oxygen atmosphere~\cite{CHO16,KARR24}. Additionally, the~interplay between defect healing and exposure to visible light in He$^+$-ion-irradiated YBCO films has been investigated~\cite{MARK10}.

In this study, we investigate the relaxation of irradiation-induced changes in the superconducting properties. Firstly, we report on the results of in situ resistance measurements during and after the irradiation process using a collimated ion beam. We also discuss the impact of sample temperature on defect formation and relaxation. Next, we investigate the thermal annealing of the irradiated YBCO films in an inert gas atmosphere and study the relaxation of defects through oxygen diffusion within the material. Finally, we report on the long-term stability of dense pinning arrays in YBCO films that were fabricated by He-FIB irradiation using a helium ion microscope (HIM). These nanostructured YBCO films show vortex matching in applied magnetic fields~\cite{AICH19,KARR24P}, and~we demonstrate the long-term stability of these matching effects together with an overall increase in the critical~current.


\section{Results and~Discussion}

\subsection{Defect Formation and Relaxation during and after He$^+$ Irradiation}

\begin{figure}[H]
\includegraphics[width=0.8\columnwidth]{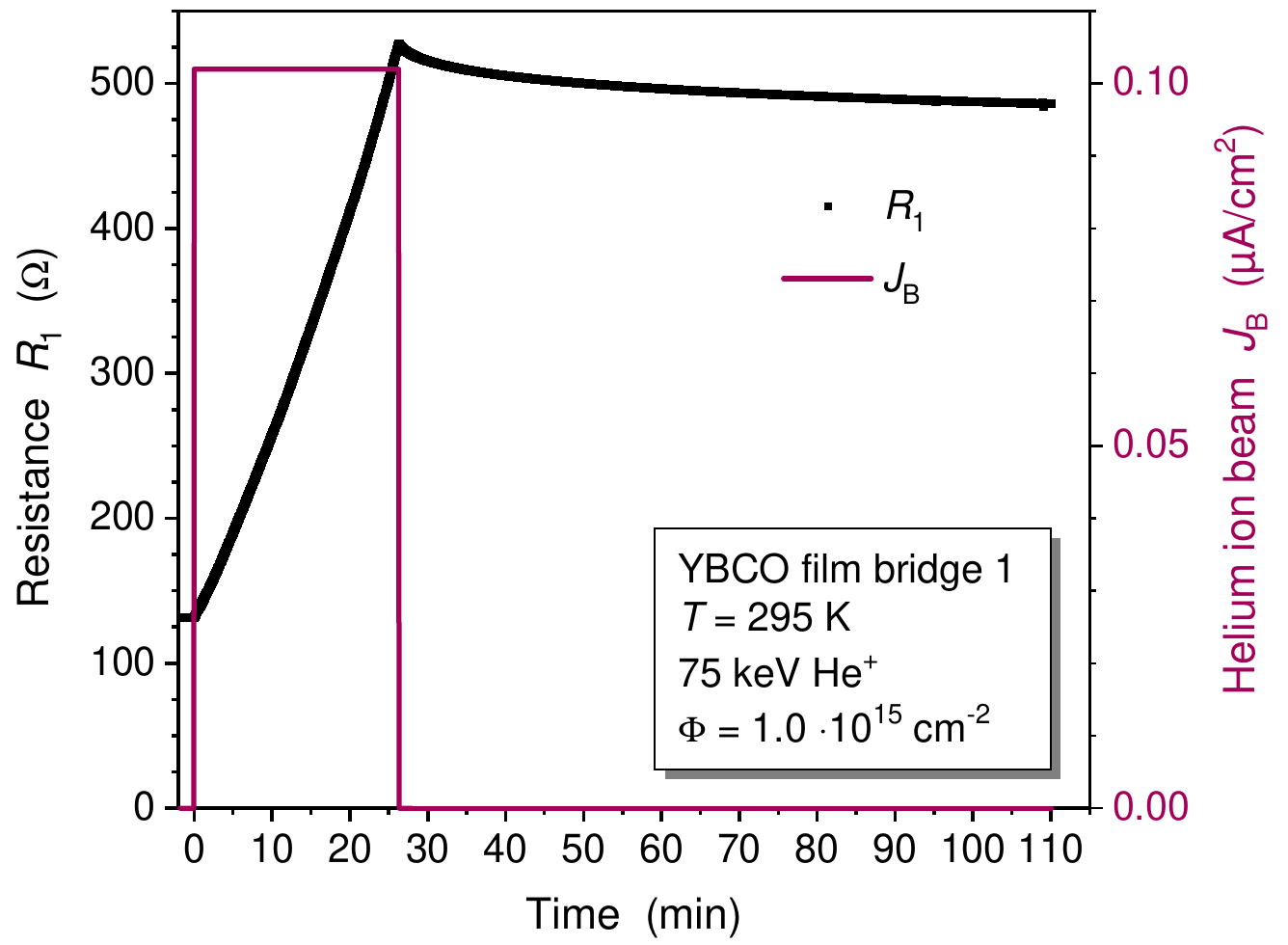}
\caption{{In-situ} resistance measurement of YBCO thin film bridge~1 during and after 75 keV He$^+$-ion irradiation at a temperature of $T = 295$ K. A~collimated He$^+$-ion beam with a current density $J_B = 0.102~\upmu$A/cm$^2$ (red line) and a fluence $\Phi = 1.0\cdot 10^{15}$ ions/cm$^2$ was used for the~irradiation.}
\label{fig:295K}
\end{figure}

In this section, we report on the modification of 240 nm thick YBCO films irradiated at different sample temperatures using a collimated He$^+$ beam. The~formation of irradiation-induced defects in the YBCO films was monitored in situ by measuring the film's electrical resistance during and after the 75 keV He$^+$ irradiation. The~critical temperature before irradiation was $T_{c0} = 85$ K, where $T_{c0}$ is the temperature at which the resistance drops below the experimental resolution. YBCO bridge~1 was irradiated at a sample temperature of $T = 295$ K with an ion-beam current density of $J_B = 0.102~\upmu$A/cm$^2$ for an irradiation time of $t_{ir1} = 26.27$ min (red line in Figure~\ref{fig:295K}). An~ion fluence $\Phi = 1.0\cdot 10^{15}$ ions/cm$^2$ was~applied.

The resistance of bridge~1, $R_1(t)$, significantly increased during He$^+$ irradiation, reaching a final value of the normalized resistance $R_1(t_{ir1})/R_1(0) = 4.01$ at time $t = t_{ir1}$ when the irradiation ended. The~resistance increase during irradiation showed a slightly superlinear behavior over time. We attribute this increase in resistance to the displacement of atoms, mainly oxygen atoms in CuO chains~\cite{ARIA03}, distorting the YBCO unit cells, while the overall oxygen content in the films remains unchanged. It is known that such an increase in resistance can occur exclusively due to the disorder of oxygen atoms without any oxygen loss~\cite{WANG95b}. After~the irradiation, the~bridge resistance slightly relaxed from its maximum value, and~it was reduced by approximately 7.8\% at time $t = 110$ min.

\begin{figure}[H]
\includegraphics[width=0.8\columnwidth]{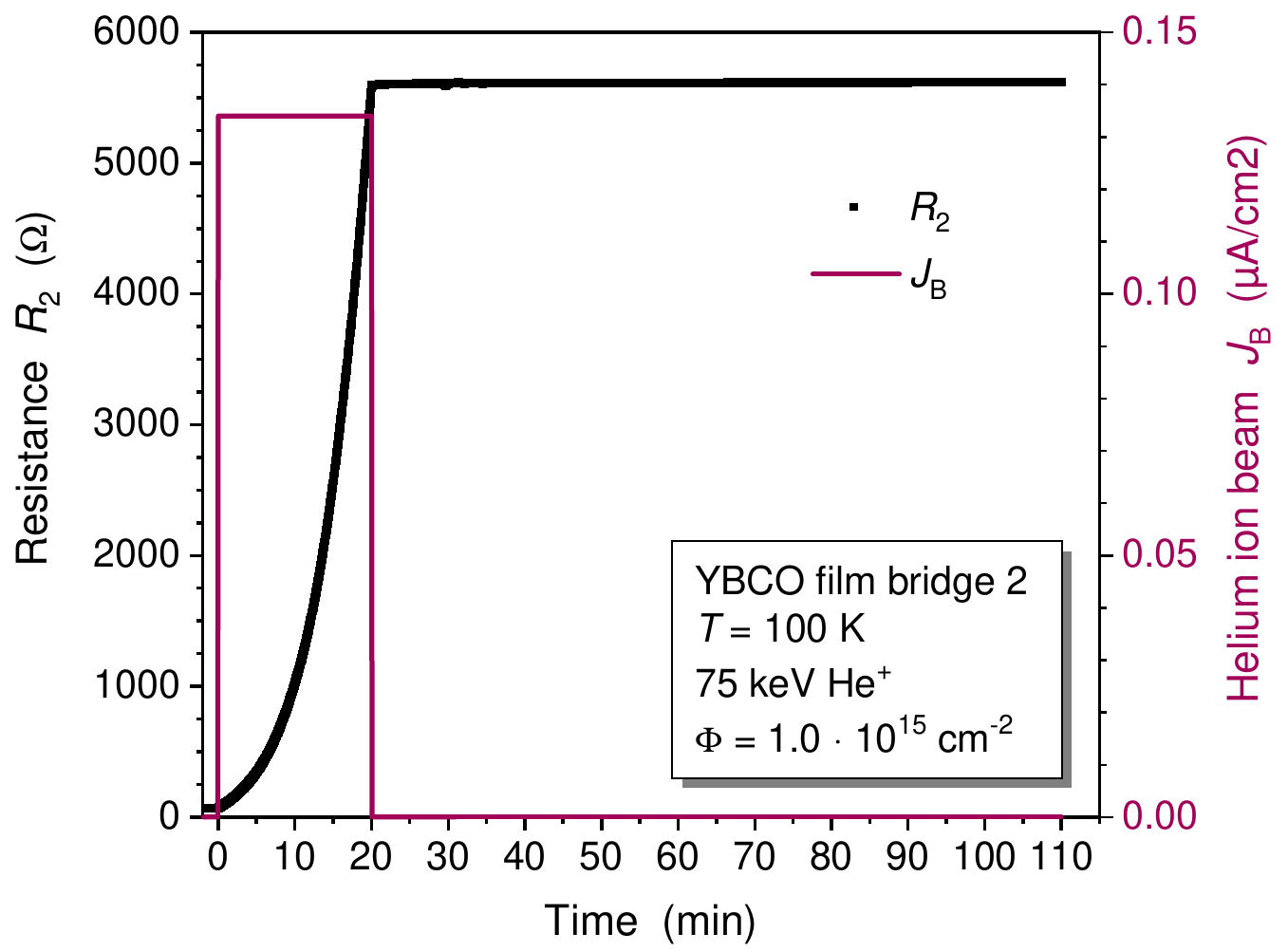}
\caption{{In-situ} resistance measurement of YBCO thin film bridge~2 during and after 75 keV He$^+$-ion irradiation at a temperature of $T = 100$ K. A~collimated He$^+$-ion beam with a current density $J_B = 0.134~\upmu$A/cm$^2$ (red line) and a fluence $\Phi = 1.0\cdot 10^{15}$ ions/cm$^2$ was used for the~irradiation.}
\label{fig:100K}
\end{figure}

In a contrasting experiment, YBCO film bridge~2 was irradiated with He$^+$ ions at a low temperature of $T = 100$ K using a cryogenic sample holder. The~sample was kept at 100 K throughout the entire measurement. The~ion current density was slightly higher at $J_B = 0.134 ~\upmu$A/cm$^2$ (red line in Figure~\ref{fig:100K}). Therefore, a~shorter irradiation time \mbox{$t_{ir2}$ = 20.02 min} was used in order to reach the same fluence of $\Phi = 1.0\cdot 10^{15}$ ions/cm$^2$.

The impact of low-temperature He$^+$ irradiation is much stronger than at room temperature. The~resistance $R_2(t)$ shows a huge and highly superlinear increase in YBCO film bridge~2. The~normalized resistance was $R_2(t_{ir2})/R_2(0) = 81.2$ at time $t = t_{ir2}$ when irradiation was stopped. In~strong contrast to the room-temperature experiment, no reduction in the bridge resistance was observed later on ($t > t_{ir2}$) when the sample was kept at a low temperature.
A related work using He$^+$-ion energies of 500 keV supports our results, as~no relaxation of resistance was found for temperatures below 250 K~\cite{BARB92}.

\unskip

\begin{figure}[H]
\includegraphics[width=0.8\columnwidth]{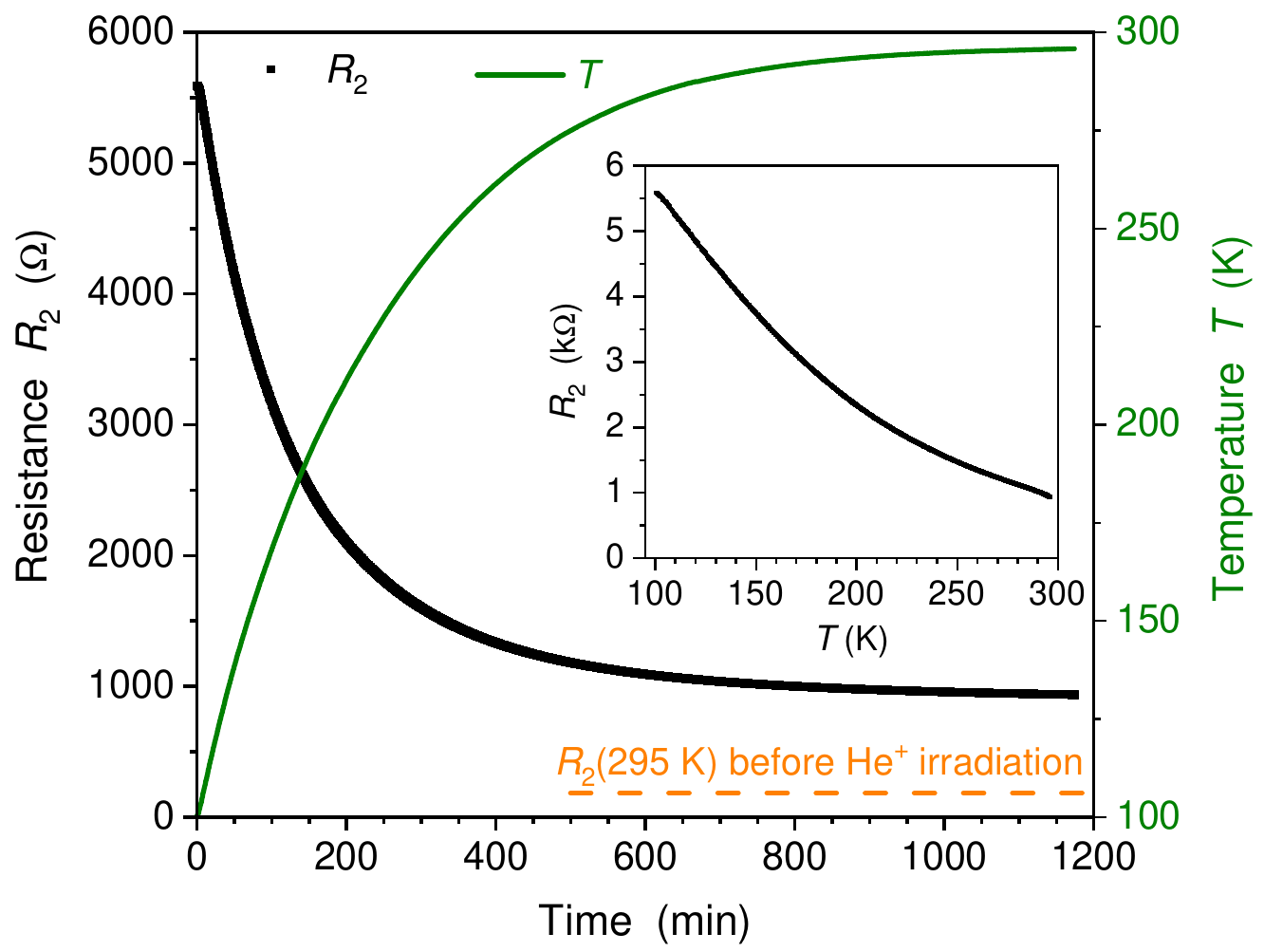}
\caption{Resistance $R_2$ and temperature $T$ of the ion-irradiated YBCO bridge~2 during warming up from 100 K to 295 K. The~room-temperature resistance before ion irradiation $R_2(295 K)$ is indicated by an orange dotted line. Inset: $R_2$ as a function of~temperature. }
\label{fig:100Kwarmup}
\end{figure}

After being exposed to ion irradiation and stored at $T = 100$ K, bridge~2 was slowly warmed up to room temperature. The~changes in resistance $R_2$ and temperature $T$ over time are shown in Figure~\ref{fig:100Kwarmup}. Remarkably, the~resistance decreased from $R_2(\text{100 K}) = 5590~\Omega$ after irradiation to $R_2(\text{295 K}) = 926~\Omega$ as the sample warmed up. The~correlation of resistance $R_2$ with temperature $T$ in the inset of Figure~\ref{fig:100Kwarmup} clearly indicates a thermally activated relaxation of irradiation-induced defects in the YBCO film. For~comparison, the~resistance of non-irradiated YBCO would have \emph{increased} by approximately a factor of three within the same temperature interval. However, the~fact that $R_2$ did not relax to the initial value before irradiation suggests that not all defects were~healed.

After the warm-up, the~resistance $R_2(295 \text{K}) = 926~\Omega$ was significantly higher than the bridge resistance before irradiation, $R_2(295 \text{K}) = 182~\Omega$, as~indicated by the dotted line in Figure~\ref{fig:100Kwarmup}. Even after a storage period of 14,000 min (almost 10 days) at room temperature, the~resistance remained elevated at $R_2(295 \text{K}) = 836~\Omega$. The~ratio of $R_2$ values measured before irradiation and after extended storage was 4.6. For~comparison, the~equivalent ratio for YBCO bridge~1, subjected to irradiation and subsequent relaxation at room temperature, was 3.7, a~value similar to that of bridge~2. These results lead to the conclusion that 75~keV~He$^+$ irradiation of YBCO thin films leads to defects with long-term stability at room temperature. Furthermore, it supports the assertion that beyond~a certain defect density, room temperature is not high enough to provide the necessary energy for the complete healing of defects following irradiation~\cite{LIU91,ILIFFE21}.

The dependence of thin film resistance on ion fluence has been modeled only very rarely in the literature. For~NiAl metal alloy films that were irradiated at $T = 77$ K with 540 keV Bi, 360 keV Xe, 120 keV Ar, and~180 keV He ions, the~observed change in film resistivity was described by a model~\cite{JAOU87,MIRA97,KARP95}. The~fraction of ion-modified material in the film was assumed to follow the Johnson–Mehl–Avrami–Kolmogorov (JMAK) equation, which describes the temporal evolution of phase transformation of solids under isothermal conditions~\cite{CHRISTIAN}. The~change in film resistivity was assumed to be proportional to this damaged fraction~\cite{JAOU87}. However, the~ion-induced change of NiAl resistivity was small and qualitatively very different from our results on YBCO films (Figures~\ref{fig:295K} and~\ref{fig:100K}), and~that model was therefore not applicable to fit our~data.

To model the increase in YBCO thin film resistance during ion irradiation, we propose another phenomenological model that takes into account the spatial distribution of irradiation-induced defects and the volume fraction of defective material in the film. When low-energy light ions impinge on the sample, they create point-like defects in the YBCO thin film. We assume that YBCO unit cells containing an irradiation defect have increased electrical resistivity $\rho + \Delta\rho$ compared to the resistivity $\rho$ of pristine unit cells. To~simplify the model, we consider that multiple defects in the same unit cell have the same impact as a single defect. The~fraction $f$ of material with irradiation defects depends on the time $t$ the sample is exposed to the ion beam, where $0 \leq f(t) \leq 1$. The~resistance of the irradiated film $R(t)$, normalized to the resistance of the pristine film $R_0$, is then given by
\begin{equation}
\frac{R(t)}{R_0}= 1 + \frac{f(t) r}{1 + r (1- f(t)^d)}.
\label{eq:j1}
\end{equation}

The normalized resistance depends on the increase in resistivity $r = \Delta\rho/\rho$, the~spatial distribution of defects described by parameter $d$, and~the fraction $f$. The~fraction of ion-modified material is expressed as $f(t) = 1-\exp(-t/\tau)$, where $\tau$ is a time constant. This ansatz corresponds to the JMAK model in its simplest form, and it takes into account that the fraction of sample material to be modified by irradiation is finite (f $\leq 1$). The~dimensionality parameter $d$ characterizes, for~instance, a~two-dimensional (2D) random distribution of columnar defects extending through the entire depth of the film ($d = 1/2$) or a three-dimensional (3D) random distribution of point-like defects in the film ($d = 2/3$). A~series and a parallel connection of two resistors would be described by $d = 0$ and $d = 1$, respectively. The~proposed model replicates Matthiessen’s rule for $d = 0$ and $f \ll 1$ and~also accounts for defect-induced modifications for larger values of $f$ while considering the dimensionality of the defect structure in the material. The~time constant $\tau$ corresponds to a characteristic ion fluence $\Phi_{ch} = J_B \tau/q$ at which the resistance experiences a notable increase. Here, $q$ is the elementary charge unit of the projectile~ion.

\begin{figure}[H]
\includegraphics[width=0.78\columnwidth]{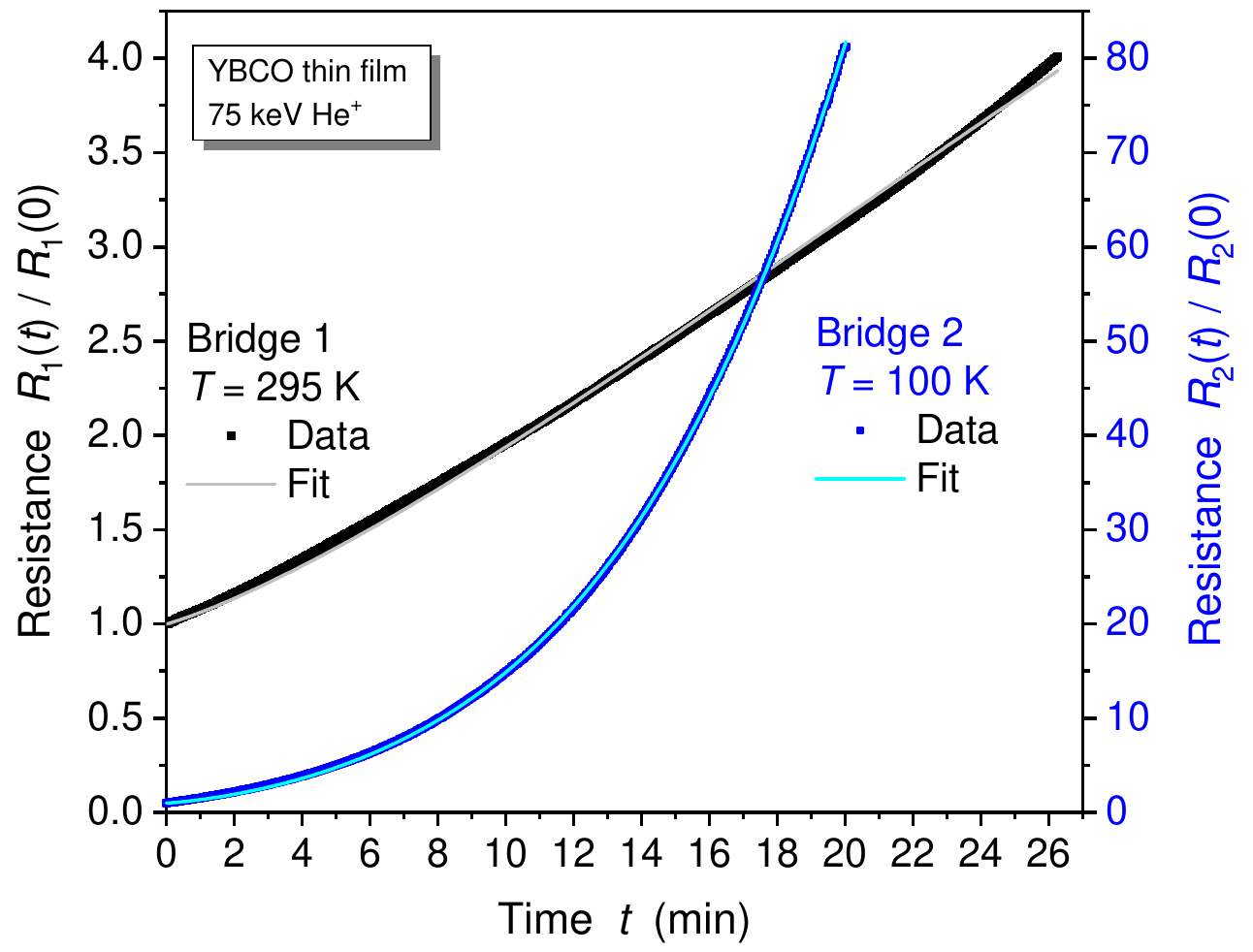}
\caption{Normalized resistance of YBCO thin film during 75 keV He$^+$ irradiation of bridge 1 at room temperature (black symbols) and bridge 2 at $T = 100$ K (blue symbols). Solid lines are fits to the data. The~dimensionality parameter extracted from the fit at $T = 100$ K was used as a fixed parameter for the fit of the data at 295 K as~well.}
\label{fig:model}
\end{figure}

The model function in Equation~\eqref{eq:j1} was fitted to the resistance of both of the irradiated YBCO thin film bridges, and~very good agreement was obtained (see Figure~\ref{fig:model}). For~film bridge 2, ion-irradiated at $T = 100$ K, the~strongly superlinear variation of resistance with time of irradiation is well described by the model function (light-blue solid line in Figure~\ref{fig:model}). The~fit parameters are given in Table~\ref{table:1}; the coefficient of determination was CoD = 99.994\%, and~the chi-square value was $\chi^2 = 0.02636$. The~spatial distribution of irradiation-induced defects was assumed to be independent of the sample temperature. Therefore, the~same value of parameter $d$ was used for the other bridge. The~resistance increase in film bridge 1, ion-irradiated at $T = 295$ K, is also well described by this model (light-gray solid line in Figure~\ref{fig:model}; $\text{CoD} = 99.894\%$, $\chi^2 = 0.00081$).

\begin{table}[H]
\centering
\caption{Time constant $\tau$, resistivity increase $r$, dimensionality parameter $d$, and~characteristic ion fluence $\Phi_{ch}$ of YBCO thin films irradiated with 75 keV He$^+$ ions at different sample temperatures. Parameters were determined from fits of the model function Equation~\eqref{eq:j1} to the normalized resistance $R(t) / R_0$ data.}
\newcolumntype{L}{>{\raggedright\arraybackslash}X}

\begin{tabularx}{\textwidth}{LLL}
\toprule

\textbf{Parameter} & \textbf{Bridge 2 (\emph{T} = 100 K)} &\textbf{ Bridge 1 (\emph{T} = 295 K)} \\  \midrule

$\tau$ (min) & 5.83 $\pm$ 0.02  & 21.60 $\pm$ 0.08 \\
$r$ & 422 $\pm$ 10  & 7.06 $\pm$ 0.06  \\
$d$ & 0.294 $\pm$ 0.003  & 0.294  \\
$\Phi_{ch}$ ($10^{15}$ cm$^{-2}$) & 0.293 $\pm$ 0.001  & 0.825 $\pm$ 0.003  \\
\bottomrule

\end{tabularx}
\label{table:1}

\end{table}

The shorter time constant and lower characteristic ion fluence, as~well as the higher resistivity increase observed for bridge~2, demonstrate a more efficient modification of YBCO films when He$^+$ irradiation is conducted at low temperature as opposed to the irradiation at room temperature of bridge~1. This is attributed to a temperature-dependent ``annealing process'' for some of the irradiation defects. The~relaxation of resistance subsequent to irradiation at \mbox{$T = 295$ K} (Figure~\ref{fig:295K}) and during warm-up after irradiation at $T = 100$ K (Figure~\ref{fig:100Kwarmup}) supports this conclusion. Despite its relaxation, the~resistance of irradiated YBCO films remains well above the intrinsic resistance prior to irradiation, persisting at elevated levels even after months of sample storage at room temperature. This shows that the irradiation process generates a significant density of stable defects. A~more sophisticated model that includes the formation of multiple defects in the same unit cell and the temperature-dependent relaxation of defects could result in even more precise fits to the $R(t) / R_0$ data obtained through~measurement.

\subsection{Defect Relaxation during Thermal~Annealing }

The results of the previous measurements unequivocally demonstrate a thermally activated or assisted defect relaxation process. Therefore, we now aim to determine the activation energy for this phenomenon. While there is extensive literature on the diffusion and the activation energy of oxygen in- and out-diffusion in YBCO~\cite{ERB96} as well as damage healing after heavy ion bombardment~\cite{MATS92b}, little is known about defect healing of point defects resulting from light-ion irradiation~\cite{ARIA03,MARK10}.

\begin{figure}[H]
\includegraphics[width=0.76\columnwidth]{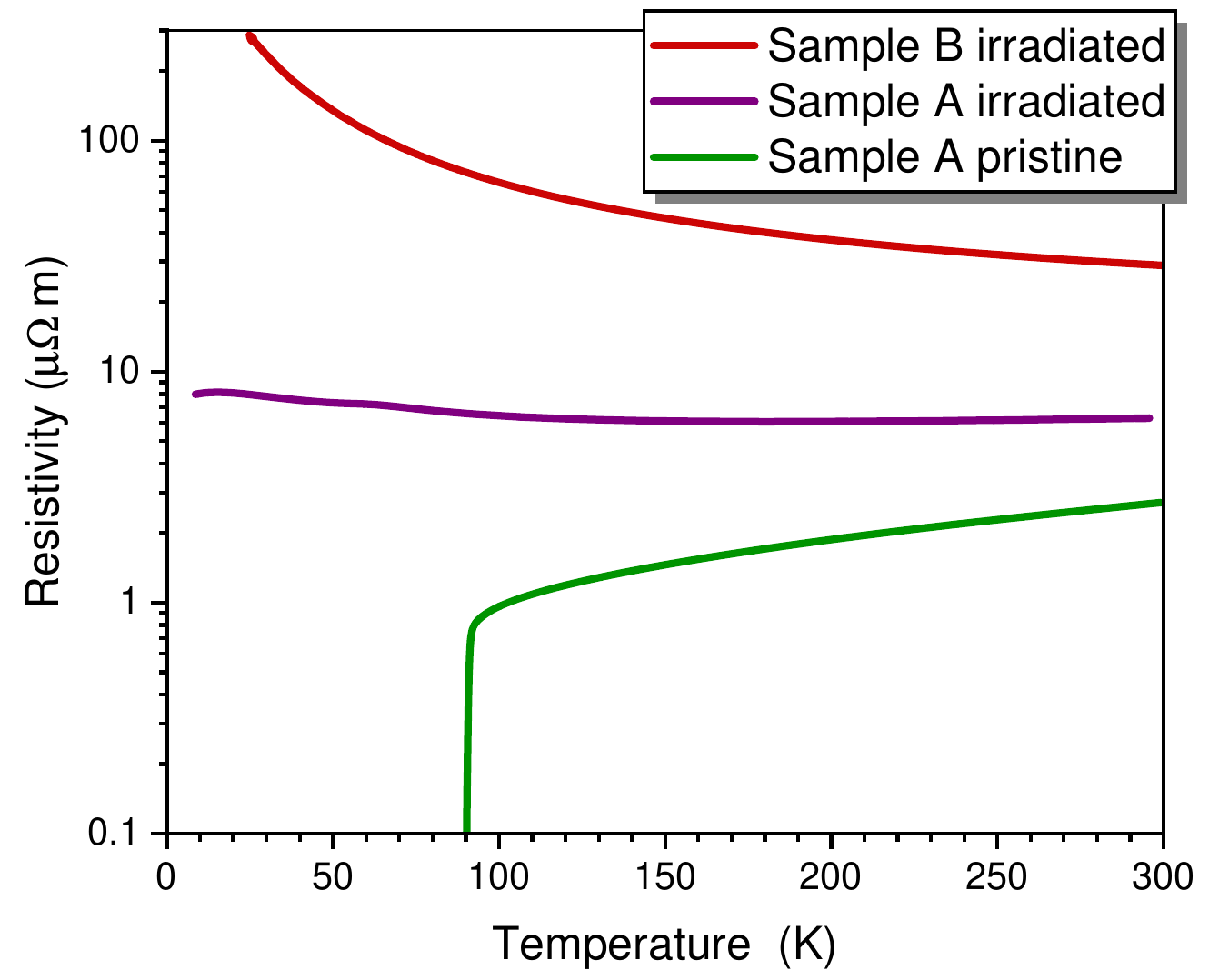}
\caption{Temperature dependence of the resistivities of thin YBCO films: sample A before (green) and after irradiation with 75 keV He$^+$ at a fluence of $0.7 \times 10^{15}$ ions/cm$^2$ (lilac) and sample B after irradiation with a fluence of $1.4 \times 10^{15}$ ions/cm$^2$ (red).}
\label{fig:MAsamples}
\end{figure}

Two 190 nm thick YBCO films were deposited on MgO and irradiated at room temperature with 75 keV He$^+$ ions with fluences of $0.7 \cdot 10^{15}$ ions/cm$^2$ (sample~A) and \mbox{$1.4 \cdot 10^{15}$}~ions/cm$^2$ (sample~B). The~temperature-dependent resistivity of the samples measured before and after ion irradiation reveals a high $T_{c0}$ of 90 K for the pristine sample and suppression of superconductivity in irradiated samples (Figure~\ref{fig:MAsamples}). After~irradiation, the~resistivity of sample~B was much higher than in sample A due to the higher irradiation~fluence.

\begin{figure}[H]
\includegraphics[width=0.85\columnwidth]{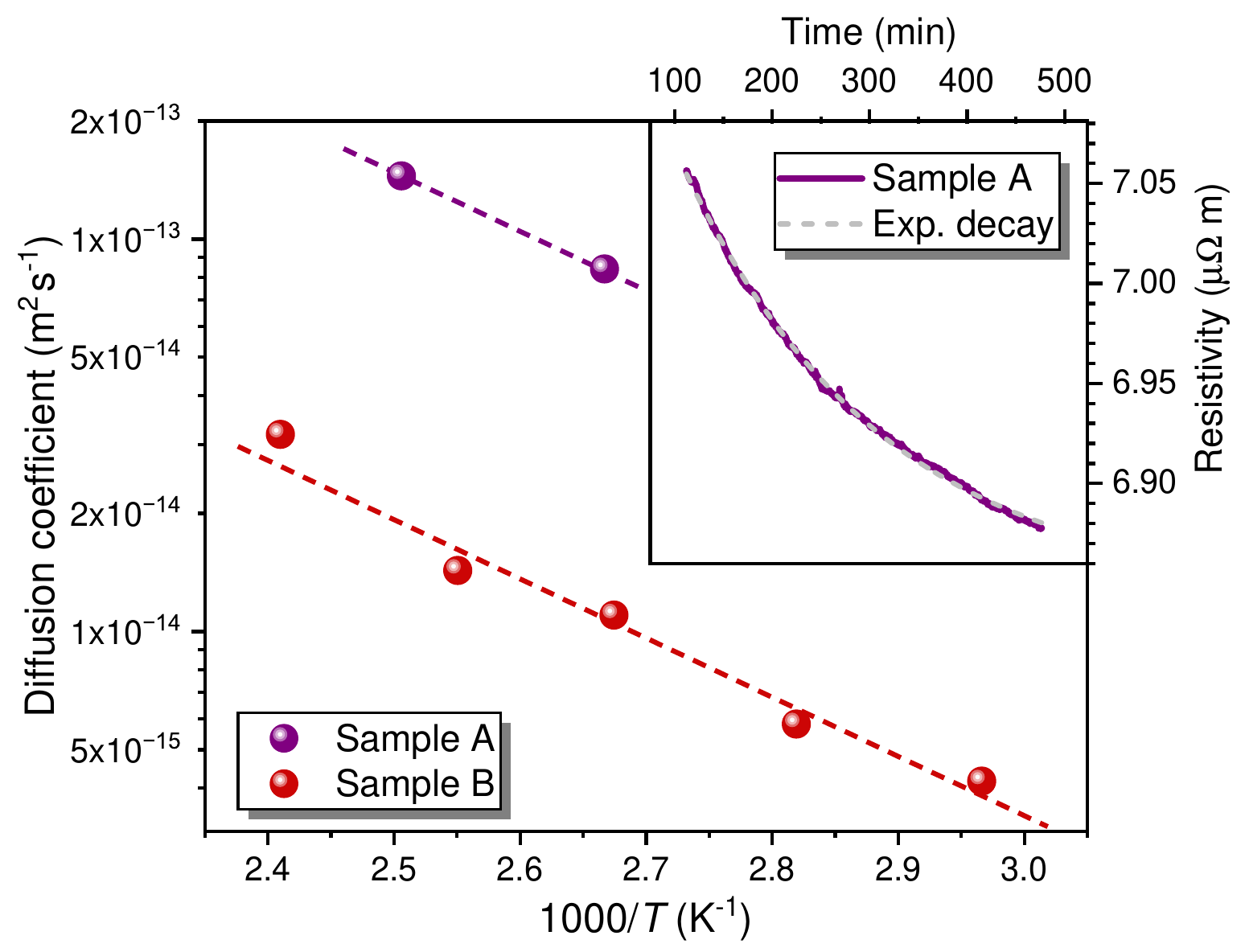}
\caption{{Diffusion$-$}
coefficient $D$ of oxygen in samples A and B after ion irradiation, determined from the exponential decay of the resistivity. Samples were kept at constant temperatures in the argon atmosphere. The~broken lines are fits to determine the activation energy for the rearrangement of oxygen atoms. The~inset shows a representative example of the resistivity decrease in sample A at $102~^\circ$C.}
\label{fig:diffusion}
\end{figure}

For the annealing experiments, the~ion-irradiated samples were placed in a quartz tube and heated in an inert Ar atmosphere while the change in resistance was recorded. Annealing temperatures were selected to be below $150~^\circ$C because previous test experiments suggested that at higher temperatures, the~measurements could be affected by oxygen loss. Reduction of the oxygen content would lead to an increase in resistivity, possibly obscuring the effects of thermally activated diffusion of the displaced oxygen atoms~\cite{MESARWI91}. After~warming up the sample to the target temperature and waiting for stable conditions, the~resistivity decay was recorded.
A representative behavior of defect annealing is depicted in the inset of Figure~\ref{fig:diffusion} for sample A at temperature $T=102~^\circ$C. The~broken gray line confirms a perfect exponential decay of the resistivity with time. Subsequently, sample A was heated to $T=126~^\circ$C and the measurement was repeated. A~similar protocol was used for sample B, encompassing more temperatures, namely, $64~^\circ$C, $81~^\circ$C, $101~^\circ$C, $119~^\circ$C, and~$142~^\circ$C.\vspace{-6pt}

\vspace{-6pt}

For further analysis, we utilize the isothermal electric resistance relaxation technique, a~common method for investigating oxygen diffusion in high-temperature superconductors~\cite{ZHANG02}. It allows us to calculate the diffusion coefficients $D$ at a given temperature $T$ by measuring the change in resistivity $\rho$ over time $t$ and subsequently determine the required activation energy $\Delta E$ in YBCO thin films that have been irradiated with He$^+$ ions.

The method is based on the relationship between oxygen concentration within the CuO$_2$ planes and the CuO chains of YBCO with resistivity. Since measuring the oxygen concentration within these CuO chains is challenging, the~changes in resistivity are assumed to be linearly correlated with the oxygen defects in the sample~\cite{DIOSA97}.

The increase in resistivity during light-ion irradiation is mostly caused by the displacement of weakly bound oxygen atoms from the copper-oxide chains. These oxygen atoms are only bound by approximately 1 eV~\cite{CUI92}, which is much lower compared to oxygen in CuO$_2$ planes (8 eV) \cite{TOLP96} or any of the other components: yttrium (25 eV), barium (30 eV), and~copper (15 eV) \cite{KIRS92}. As~a result, most irradiation-induced displacements are oxygen atoms 
from the chains~\cite{CHO16} that are displaced by a small distance from their original position while not decreasing the oxygen content in the sample~\cite{ARIA03}. It is commonly assumed that the displaced oxygen species are often lodged along the $a$ axis of the basal plane of the YBCO unit cell~\cite{VALLES89}, where minimal energy is needed to shift them back into place again~\cite{ARIA03}.

Annealing can be used to let the displaced oxygen diffuse back into place, thereby lowering the resistivity. It is important to note that this diffusion is a thermally activated process, and~the changes in resistivity over time at a fixed annealing temperature need to be recorded. Since oxygen diffusion perpendicular to the CuO$_2$ layers in YBCO is negligible, predominant diffusion within the $ab$ plane can be assumed~\cite{ERB96}.

Fick's second law can be used to describe this time-dependent diffusion process, which~states
\begin{equation}
\frac{\partial c}{\partial t} = D \frac{\partial^2 c}{\partial^2 x}.
\label{eq:1}
\end{equation}

In this scenario, we examine those oxygen atoms in YBCO which form the chains along the $b$ axis. We consider their concentration $c$ as a function of time $t$ and \mbox{position $x$}. The~diffusion coefficient $D$ is defined by an Arrhenius equation that depends on the temperature $T$ and the activation energy $\Delta E$:
\begin{equation}
D(T)= D_0 \cdot \exp \left(- \frac{\Delta E}{k_B T} \right) ,
\label{eq:2}
\end{equation}
where $D_0$ is a material constant and $k_B$ is the Boltzmann~constant.

When solving Equation \eqref{eq:1}, we assume as the initial condition a homogeneous distribution of oxygen in the CuO chains of the sample and limit the process to the sample volume. This allows us to find a solution for the concentration. Subsequently, we can determine the relative changes in concentration~\cite{ERB96}.
\begin{equation}
\frac{c (t)-c_e}{c_0-c_e} = \frac{8}{\pi^2} \cdot \exp \left(- \frac{t}{\tau} \right)
\label{eq:3}
\end{equation}
with \[\tau = \frac{b^2}{\pi^2 D} ,\]
where $c_e$ represents the saturation concentration, indicating the concentration of all the oxygen atoms positioned correctly in the CuO chains; $c_0$ is the starting concentration, and~$c(t)$ is the concentration over time. The~relaxation time $\tau$ depends on the sample width $b$ and diffusion coefficient $D$.

Because measuring the concentration of oxygen in CuO-chain positions is hardly possible, we used the linear relationship between the concentration of displaced oxygen atoms and resistivity from measurements at various ion fluences~\cite{ARIA03}. For~temperatures low enough so only small changes in oxygen occur, Equation~\eqref{eq:3} can be converted to
\begin{equation}
\ln \left[ \frac{\rho (t)-\rho_e}{\rho_0-\rho_e} \frac{\pi^2}{8} \right] = - \frac{t}{\tau} ,
\label{eq:4}
\end{equation}
where $\rho (t)$ is the time-dependent resistivity, $\rho_0$ is the initial resistivity of the irradiated sample before annealing, and~$\rho_e$ is the saturation resistivity for $t \rightarrow \infty$ . For~more details on the derivation of Equation~\eqref{eq:4}, see~\cite{ERB96}.

In order to evaluate the oxygen diffusion coefficient $D(T)$ in the irradiated samples, we measured $\rho(t)$ during the annealing at temperature $T$ and utilized Equation~\eqref{eq:4}. Since $\rho_e$ is not accessible experimentally, the~measurements were performed over timescales significantly larger than $\tau$ and $\rho_e$ determined from a fit of the exponential decay of $\rho(t)$ as exemplified in the inset of Figure~\ref{fig:diffusion}. The~values of $D$ at several different temperatures are displayed as an Arrhenius plot in Figure~\ref{fig:diffusion}. The~slope of the fits (broken lines) allows us to determine the activation energy $\Delta E$ using Equation~\eqref{eq:2}. Note, however, that the absolute values of $D$ are subject to assumptions in the model that enter the prefactor $D_0$, which has no influence on the determination of $\Delta E$.

The results are $\Delta E_{B} = (0.31 \pm 0.03)$ eV for sample B and $\Delta E_{A} = 0.29$ eV for sample A. It is important to note that the value for sample A is based on only two measurements. Nonetheless, it still indicates that the activation energy is similar for different irradiation fluences.
Other oxygen-diffusion experiments reported somewhat higher values for the activation energy, e.g.,~1.23 eV~\cite{CHEN92}, 0.8 eV~\cite{ERB96}, and~0.97 eV~\cite{ROTH91}. However, these experiments were conceptually different, as~they measured the in- and out-diffusion of oxygen during annealing experiments in an oxygen atmosphere. In~this scenario, diffusion is influenced by surface barriers~\cite{TU89}, while in our experiments, the~average oxygen content in the sample remains constant. A~better-related study of defect recovery after irradiation with 500 keV He$^+$ ions corroborates our findings and found a value of $\Delta E = (0.36 \pm 0.05)$ eV~\cite{BARB92}.

\subsection{Long-Term Stability of Nanopatterned YBCO~Films}

Nanopatterning of YBCO thin films by focused-ion-beam irradiation has proven to be a versatile method for fabricating ultradense vortex pinning landscapes and Josephson junctions. However, one might have concerns about their long-term stability due to oxygen migration that could blur the oxygen-defect profile~\cite{ZALU24}. Building on our previously presented experiments, we investigate the temporal evolution of the critical temperature, the~resistivity, and~the critical current of an 80 nm thick YBCO film on MgO substrate that was irradiated with a 30 keV He$^+$ focused ion beam in the HIM at room temperature. The~columnar defects (CDs) were arranged in a square lattice with 200 nm spacings. More details on sample fabrication and properties are reported elsewhere~\cite{AICH19}.

\begin{figure}[H]
\includegraphics[width=0.8\columnwidth]{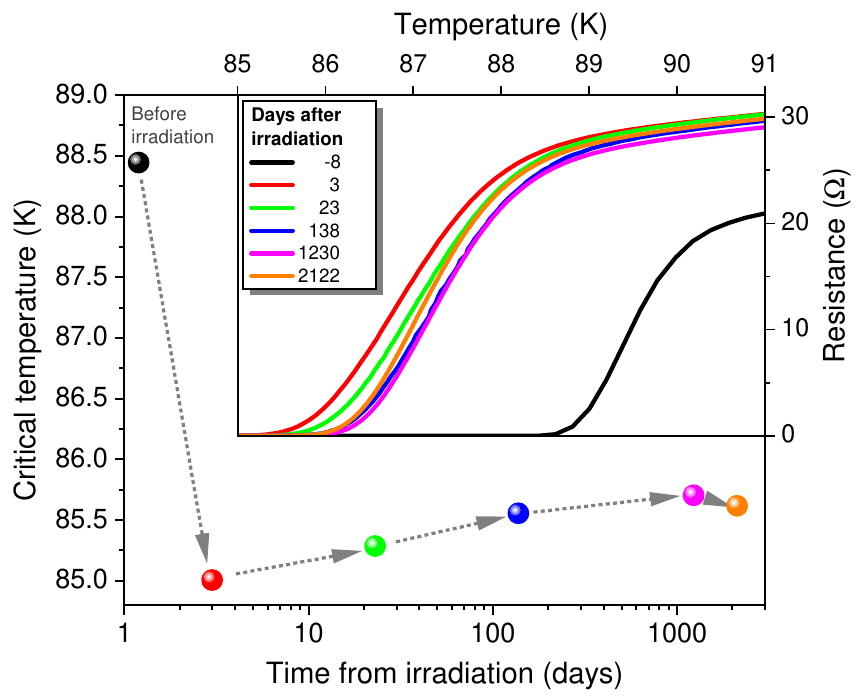}
\caption{{Zero-resistance} critical temperature $T_{c0}$ over room-temperature storage time in an 80 nm YBCO film before and after He-FIB irradiation imprinting a square pinning array of 200 nm spacings. The~black bullet represents $T_{c0}$ eight days before the irradiation. The~arrows indicate the sequence of $\rho(T)$ measurements from which $T_{c0}$ was determined, using a 10 m$\Omega$ criterion. Inset: Evolution of the resistance vs. temperature characteristics of the sample with room-temperature annealing time. The~black line represents the $\rho(T)$ measurement of the sample before~irradiation.}
\label{fig:RT_S200}
\end{figure}

The temporal evolution of the resistivity vs. temperature characteristics and the change in $T_{c0}$ are illustrated in Figure~\ref{fig:RT_S200} over a span of almost six years. The~sample was kept in a desiccator at room temperature in the ambient atmosphere between measurements. It is important to emphasize that in our analysis we are only considering the days when the sample was stored at room temperature, as~aging effects are minimal when the sample is kept in the cryostat at low temperature. After~irradiation, the~critical temperature decreased by approximately 3.5 K due to the inevitable straggle of some ions off their path, causing a minor amount of irradiation defects between the CDs. Over~the storage period, $T_{c0}$ initially increased and reached its maximum after approximately 3.4 years. For~even longer storage periods, a~marginal decrease in $T_{c0}$ was found, accompanied by a small increase in the normal-state~resistivity.

The long-term stability of unirradiated thin YBCO films has long been a topic of controversial discussion, especially with regard to effects of exposure to humidity and carbon dioxide in air~\cite{GHERA90}. Exposure to air causes YBCO to form hydroxide and carbonate layers on its surface~\cite{BEHN93}. Additionally, the~commonly used photolithographic patterning exposes the $ab$ planes of YBCO to the environment, which are more prone to environmental degradation~\cite{BEHN93}.

However, for~irradiated samples, a~competition between defect healing after irradiation, which increases $T_{c0}$, and~the above-mentioned deterioration of the sample through decomposition, which decreases $T_{c0}$, can be anticipated. In~fact, we observe an increase in $T_{c0}$ and a reduction in resistivity during storage in a desiccator up to a time span of approximately 3.4 years, which we attribute to a partial rearrangement of displaced oxygen atoms. The~minor reversal of this trend at still longer times is probably connected with a saturation of the healing process and a small deterioration of the sample properties during dry-air exposure. Nevertheless, our data indicate a robust long-term stability of the nanopatterned YBCO films if kept in dry~air.

\begin{figure}[H]
\includegraphics[width=0.8\columnwidth]{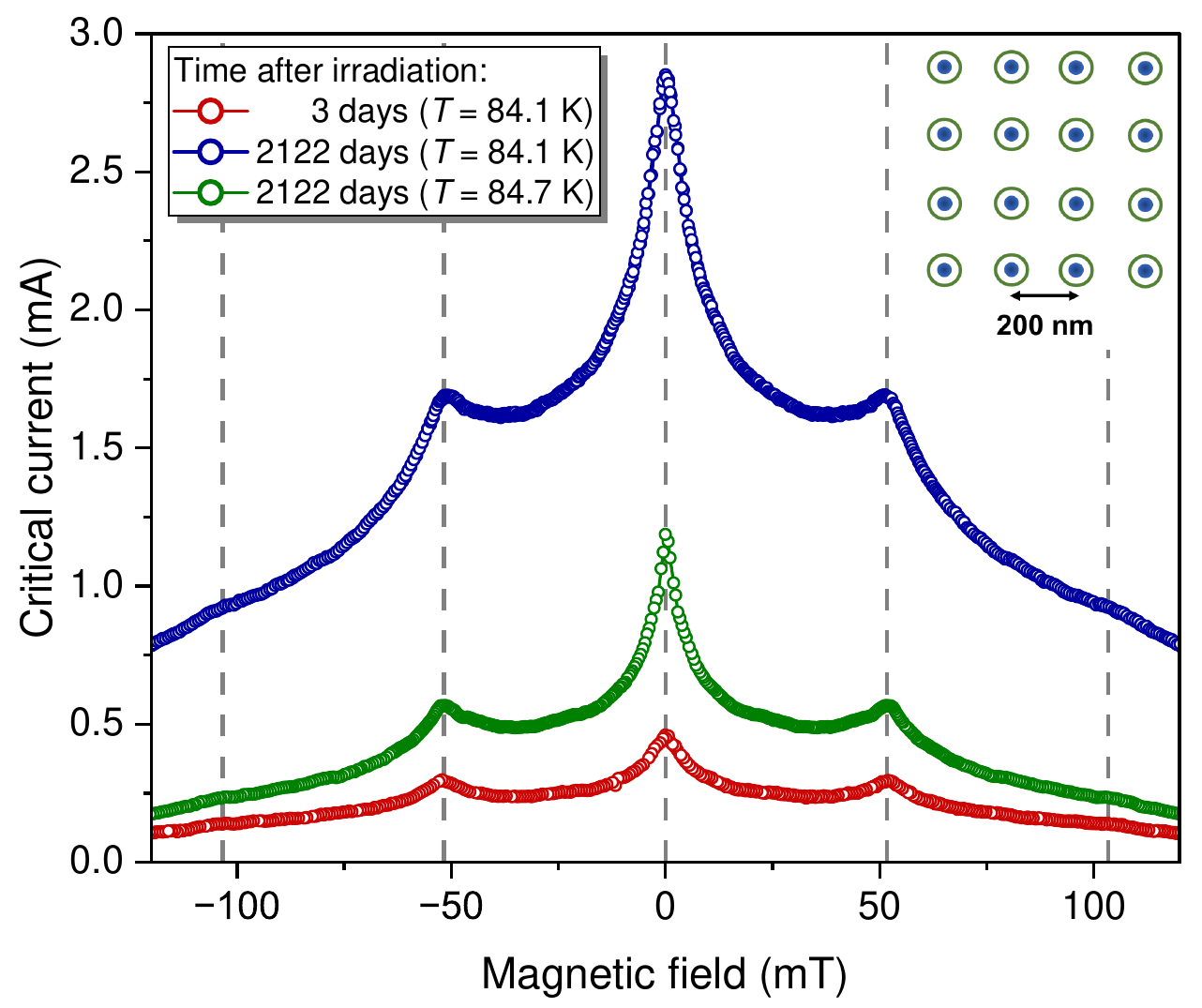}
\caption{Temporal evolution of the critical current in an 80 nm YBCO film structured by He-FIB into a square array of defect columns with 200 nm spacings. The~initial critical current measurement after irradiation is denoted by red circles, while the subsequent measurement following 2122 days (almost six years) of storage in dry air at room temperature is represented by blue circles. The~green symbols show the corresponding measurement at 84.7 K, taken at the same reduced temperature $T/T_{c0}=0.989$ as the measurement before long-term storage. Gray broken lines represent the magnetic fields according to Equation~\eqref{eq:matching} with values $k=\{-2,-1,0,1,2\}$. Inset: Layout of the pinning lattice, where green circles represent the columnar defects and blue dots denote the positions of the trapped vortices for $k=\pm 1$.}
\label{fig:matching}
\end{figure}

Vortex commensurability effects occur when a magnetic field applied parallel to the CDs creates a vortex density that matches the sample's CD arrangement. Maxima of the critical current appear at so-called matching fields
\begin{equation}
B_k = k \Phi _0/A,
\label{eq:matching}
\end{equation}
where $k$ is the number of pinning sites (or vortices) in the unit cell of area $A$ of the two-dimensional CD lattice, and~$\Phi_0=h/(2e)$ is the magnetic flux quantum. The~square array of CDs in our nanopatterned sample features a unit cell of $A=(200~\text{nm})^2$, leading to a first matching field $B_1 = 51.7$ mT~\cite{AICH19}. A~sketch of the pinning lattice (green circles) 
and the positions of the trapped vortices (blue dots) at the first matching field is shown in the inset of Figure~\ref{fig:matching}.

The sample was kept at room temperature for three days after irradiation in the HIM before~we conducted an initial measurement of the critical current in the superconducting state at $T = 84.1$ K (red circles in Figure~\ref{fig:matching}). Another measurement was performed 2122 days later under the same conditions (blue circles). The~matching field is identical, indicating that the pinning landscape is still efficient and leads to the same vortex arrangement. However, the~overall critical current has increased over the entire magnetic field range. In~particular, its value at $B_1$ became 5.8 times larger. At~first glance, this could be due to the decrease in the relative temperature $T/T_{c0}$ as a result of the higher $T_{c0}$, as~shown in Figure~\ref{fig:RT_S200}. Typically, the~critical current significantly increases at temperatures further below $T_{c0}$. However, upon~closer inspection of an additional measurement taken at the same $T/T_{c0}=0.989$ as the initial measurement, it is revealed that the critical current is still enhanced by a factor of 1.9 after long-term~storage.

The long-term resilience of the matching peaks can be understood as follows: The spreading of the penetrating He$^+$ ions causes a ``damage gradient'' perpendicular to its initial trajectory. The~density of defects in the center of the defect column is high enough to withstand room-temperature annealing. However, the~defect density rapidly decreases with the distance from the center~\cite{KARR24P}. These inter-CD regions are primarily susceptible to healing effects which potentially can ``sharpen'' the defect profile around the CDs due to Frenkel defect recombination~\cite{SIRE09} and ultimately lead to the observed increase in $T_{c0}$, the~reduction of the resistivity, and the increase in the critical current at the matching~field.

Indeed, studies on He-FIB-created Josephson junctions (JJs) found that aging at room temperature in a nitrogen environment resulted in a significant enhancement of the critical current, with~timescales that are highly dependent on the irradiation dose. Annealing at $90~^\circ$C under low and high oxygen pressures confirmed that the modifications in the JJs' 
characteristics stem from repositioning oxygen atoms to their original sites rather than restoring the oxygen content in the films~\cite{KARR24}. This discovery rules out the possibility of oxygen depletion during He$^+$-ion irradiation, aligning with electrical transport measurements~\cite{KARR24P}. Other works reported good temporal stability of JJs prepared by masked Ne$^+$-ion irradiation over eight years~\cite{CYBA13} and investigated the voltage modulation of current-biased JJ arrays as a function of the applied magnetic field. The~increase in the voltage modulation after post-annealing at $100~^\circ$C in oxygen resulted from a narrowing of the barrier by diffusion and recombination of the low-energy oxygen defects~\cite{CHO16}.

The findings of these various experiments indicate that controlled annealing at temperatures moderately above room temperature could be a tool to enhance the properties of nanostructures created by He-FIB in a reasonably short timescale while also anticipating healing effects and thus increasing their long-term~stability.



\section{Methods and~Materials}

\subsection{YBCO Thin Film~Production}
{All YBCO} thin films in this work were produced by pulsed laser deposition (PLD)~\cite{BAUERLE}.
Films were grown on single-crystal MgO (100) substrates. A~stoichiometric YBCO ceramic fabricated via the solid-state reaction method was used as the target. Ablation was carried out with a KrF-excimer laser ($\lambda$~=~248 nm) at a pulse repetition rate of 10 Hz with a pulse duration of 25 ns and a fluence of 3.25 J/cm$^2$. The~deposition took place at a substrate temperature of 750~°C in a 0.7 mbar oxygen background. Annealing at 450~°C in 800~mbar oxygen background was carried out for 30 min. Finally, the~films were patterned by wet-chemical etching via a photolithographic mask to define geometries suitable for four-point resistance~measurements.

\subsection{Irradiation of Thin~Films}

The irradiation of YBCO using a collimated He$^+$-ion beam was performed on an ion implanter ({High Voltage Engineering Europa B.V., Amersfoort, the Netherlands}), using a metal mask to shield the electrical contacts of the sample. The~setup allowed for cooling of the sample during irradiation by a flow of liquid nitrogen while also measuring the resistance. He$^+$ ions with 75 keV energy were used, with~99\% of them exiting the YBCO film without being implanted at this energy level. To~prevent channeling of He$^+$ ions along the $c$ axis of the sample, the~ion incidence angle was set to $5^\circ$ off the surface normal. The~ion fluence was monitored by Faraday cups, and the irradiation was halted once the desired, preset fluence was achieved. The~ion beam current density was kept below 0.25 $\upmu$A/cm$^2$ to avoid inducing thermal effects during~irradiation.

The irradiated nanostructures were prepared using a HIM ({ORION NanoFab, Carl Zeiss Microscopy, Oberkochen, Germany}). An~array of irradiated spots with a diameter of $\approx$50 nm was created with a slightly defocused 30 keV He$^+$-ion beam at a beam current of 3.0 pA. A~total of 50,000 ions per defect column were used to form the pinning array. More detailed information on creating the array of columnar defects via HIM and comprehensive measurements on the nanoirradiated samples can be found in~\cite{AICH19}.

\subsection{Annealing of Thin~Films}

To anneal the samples above room temperature, they were placed in a quartz-glass tube inside a tube furnace ({Heraeus RO 4/25, Hanau, Germany}) with a temperature controller. A~minimal overpressure of flowing Ar gas, controlled by a gas-washing bottle at the back end of the quartz tube, was maintained. We made sure that the gas flow did not compromise the temperature stability, and~the sample temperature was continuously monitored by a platinum ({Pt-100}) sensor.

\subsection{Electrical Characterization of Thin~Films}

The in situ resistivity measurements of the films during and after irradiation were performed with a constant current supply ({Keithley Instruments 224, Solon, OH, USA}), with~a measurement current of 1 $\upmu$A, and~a nanovoltmeter ({Keithley Instruments SDV 182,  Solon, OH, USA}).

The setup for the determination of the oxygen diffusion coefficient and the activation energy consisted of a different current supply ({Keithley Instruments 6221, Solon, OH, USA)}) and nanovoltmeter ({Keithley Instruments 2182A, Solon, OH, USA}). Here, a~measurement current of 10~$\upmu$A was~chosen.

Electrical transport measurements in magnetic fields were conducted on the HIM-structured samples using a closed-cycle cryocooler positioned between the poles of an electromagnet. The~setup allows for precise temperature adjustments with a stability of approximately 1 mK. This was achieved by employing a ceramic temperature sensor (Cernox-1080, Lake Shore Cryotronics, Woburn, MA, USA) in conjunction with a temperature controller ({LakeShore Cryotronics 336, Woburn, MA, USA}). Critical currents were determined using a 100 nV criterium, corresponding to an electrical field of $10~\upmu$V/cm.


\section{Conclusions}

In summary, we examined the long-term stability and the annealing of defects created by He$^+$-ion irradiation in thin YBCO films in in situ and ex situ experiments. The~increase in resistivity depends strongly on the temperature at which the sample is kept during irradiation and can be described by a phenomenological model; while a decrease in resistivity occurs within minutes after stopping the irradiation at room temperature, no such relaxation is seen at 100 K. The~counteracting creation and healing of oxygen displacements is, therefore, an~important consideration for the chosen irradiation~temperature.

Monitoring the resistivity evolution at several elevated temperatures, where oxygen out-diffusion can still be ruled out, allowed us to determine the diffusion coefficients for rearranging oxygen atoms into their original sites. The~activation energy \mbox{$\Delta E = (0.31 \pm 0.03)$ eV} of this process is lower than the one for oxygen diffusion out of the YBCO film, making it possible to improve sample properties by annealing at selected temperatures without deoxygenating~them.

Since focused He$^+$-ion-beam irradiation is an extremely versatile tool to modulate the superconducting properties, we investigated the aging of vortex-pinning arrays over long timescales. During~the initial healing period, $T_c$ slightly increased during 3.4~years and only marginally decreased thereafter. The~general reduction of resistivity did not compromise the vortex-matching signatures. On~the contrary, we observed a dramatic increase in the critical current in the whole range of magnetic fields under investigation that is most notable at zero field and at the matching peaks. Our analysis suggests that long-term room-temperature annealing maintains the robust pinning potential at the cores of the columnar defect channels while reducing the defect density surrounding them. These results indicate the long-term stability of He$^+$-ion-irradiated YBCO films at room temperature following an initial enhancement during a healing phase, and~they underscore the robustness of structures fabricated using a focused He$^+$-ion beam. This has significant implications for potential applications in~fluxonics.

\vspace{6pt}

\authorcontributions{{Conceptualization,} J.D.P., D.K. and W.L.; methodology, B.A., B.M., J.D.P., D.K. and W.L.; validation, S.K., J.D.P., B.A. and W.L.; formal analysis, S.K., B.A., P.R., M.-A.B., J.D.P. and W.L.; investigation, B.A., B.M., P.R., M.K. and M.-A.B.; resources, J.D.P, R.K., E.G., D.K. and W.L.; data curation, J.D.P. and W.L.; writing---original draft preparation, S.K.; writing---review and editing, J.D.P. and W.L.; visualization, S.K. and B.A.; supervision, J.D.P., R.K., D.K. and W.L.; project administration, J.D.P., E.G., D.K. and W.L.; funding acquisition, J.D.P., D.K. and W.L. All authors have read and agreed to the published version of the manuscript.}
\newpage
\funding{{The} research was funded in whole, or~in part, by~the Austrian Science Fund (FWF) grant I4865-N and the German Research Foundation (DFG), grant KO~1303/16-1. For~the purpose of open access, the~authors have applied a CC-BY public copyright license to any Author Accepted Manuscript version arising from this submission. The~research is based upon work from COST Actions CA21144 (SuperQuMap), CA19140 (FIT4NANO), and~CA19108 (Hi-SCALE) supported by COST (European Cooperation in Science and Technology).}
\institutionalreview{Not applicable.} 

\informedconsent{{Not applicable.}}

\dataavailability{Data are available on reasonable request from the \mbox{corresponding author}.}

\acknowledgments{{Open Access Funding by the University of Vienna.}}

\conflictsofinterest{Author Marius-Aurel Bodea is employed by the Infineon Technologies Austria AG. All authors declare that the research was conducted at universities in the absence of any commercial or financial relationships that could be construed as a potential conflict of interest.} 

\begin{adjustwidth}{-\extralength}{0cm}

\reftitle{References}

\PublishersNote{}
\end{adjustwidth}
\end{document}